\documentclass[aps,pr,reprint,longbibliography]{revtex4-1}

\usepackage{amsmath}
\usepackage{graphicx}
\graphicspath{{figures/}}
\usepackage{siunitx}

\DeclareMathOperator{\sinc}{sinc}
\renewcommand{\vec}[1]{\mathbf{#1}}
\newcommand{\avg}[1]{\left< #1 \right>}

\newcommand{\abs}[1]{\left\vert #1 \right\vert}

\setcitestyle{square,numbers,comma}

\begin{document}

\title{Dynamics
of an acoustically trapped sphere in beating sound waves}

\author{Mohammed A. Abdelaziz}
\author{David G. Grier}
\affiliation{Department of Physics and
  Center for Soft Matter Research,
  New York University, New York, NY 10003}

\date{\today}

\begin{abstract}
A focused acoustic standing wave creates
a Hookean potential well for
a small sphere and can levitate it
stably against gravity.
Exposing the trapped sphere to a second
transverse travelling sound wave imposes an 
additional acoustical force that drives
the sphere away from its mechanical
equilibrium.
The driving force is shaped by
interference between the standing 
trapping wave and the traveling
driving.
If, furthermore, the traveling wave
is detuned from the standing wave,
the driving force oscillates at the
difference frequency.
Far from behaving like a textbook
driven harmonic oscillator,
however, the wave-driven harmonic oscillator
instead exhibits a remarkably
rich variety of dynamical behaviors
arising from the spatial dependence
of the driving force.
These include oscillations at both
harmonics and subharmonics of the
driving frequency, period-doubling
routes to chaos and Fibonacci
cascades.
This model system therefore illustrates
opportunities for dynamic acoustical
manipulation based on spectral control
of the sound field, rather than
spatial control.
\end{abstract}

\maketitle

Acoustical manipulation is emerging as an
attractive alternative to optical manipulation
for applications where large forces are
required to move sizeable objects over
macroscopic distances \cite{inoue2019acoustical,marzo2019holographic,melde2016holograms,kozuka2007noncontact,wu1991acoustical}.
Sound waves can exert far more force
per watt than light waves \cite{baresch2016observation,thalhammer2011combined}, but are not
yet so easy to control.
Optical techniques have the advantage
in this regard thanks to a wealth of technology
for controlling the spatial structure of laser 
beams.
Dynamic holographic optical trapping,
for example,
uses megapixel arrays of phase-shifting
elements to sculpt an ordinary laser beam
into complex three-dimensional optical
force fields \cite{grier2003revolution,leach20043d}.
Analogous devices for
sound are under development 
\cite{marzo2015holographic,marzo2019holographic,hirayama2019volumetric,fushimi2020ultimate},
but do not
yet offer the same degree of
control afforded by their optical 
counterparts.

One respect in which sound has a clear
advantage over light is the ease with
which a coherent sound wave's frequency can
be altered.
The present study illustrates the opportunities
for dexterous and dynamic acoustical manipulation
afforded by spectral engineering through
experimental and numerical studies of 
a deceptively simple
dynamical system assembled and actuated by two
continuous sound waves.
Slightly detuning the two waves gives rise
to a remarkably rich and potentially useful
spatiotemporal phenomenology
without requiring active control of the
force field's structure.

\begin{figure}
    \centering
    \includegraphics[width=\columnwidth]{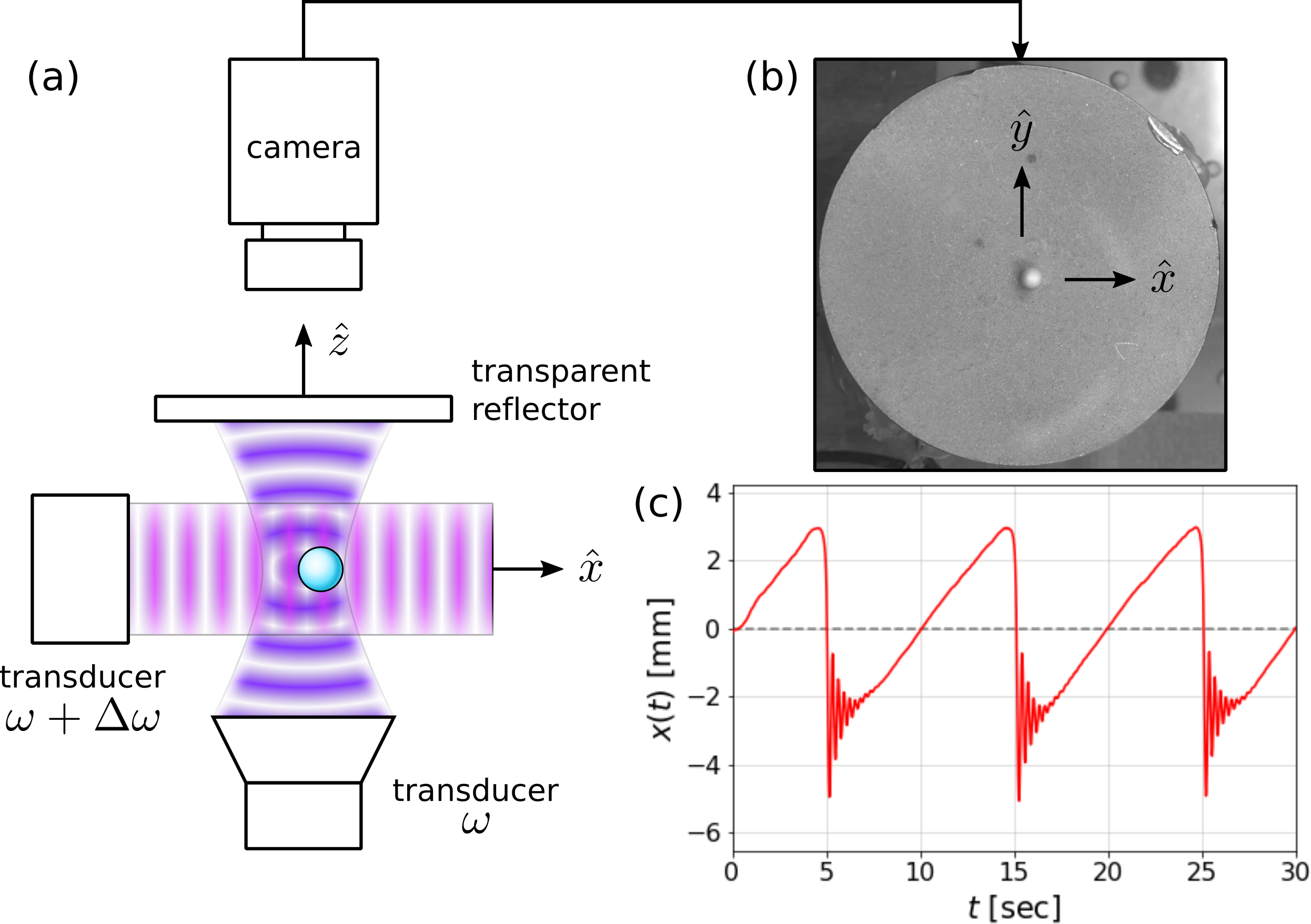}
    \caption{(a) Schematic representation of an
    acoustically levitated particle driven
    by a harmonic traveling wave.
    (b) Video frame showing a top view of the particle at
    $z = \SI{4}{\mm}$ above the transducer in
    a standing wave at $f = \SI{40.3}{\kilo\hertz}$.
    (c) Measured trajectory, $x(t)$, 
    of the particle driven by a difference
    frequency, $\Delta f = \SI{0.1}{\hertz}$.}
    \label{fig:schematic}
\end{figure}

Our system, shown schematically
in Fig.~\ref{fig:schematic}(a),
consists of a small sphere
of mass $m$
that is stably levitated in air by a
vertical standing-wave acoustical trap
at frequency $\omega$.
The standing wave's transverse
intensity gradients give rise
to a restoring force that localizes
the sphere in a three-dimensional harmonic well with
natural frequency $\omega_0 \ll \omega$.
The trapped sphere also is exposed to a
travelling sound wave at
frequency $\omega + \Delta \omega$
that is projected horizontally.
Interference between the two sound waves
creates a force field
for the sphere that oscillates at
the difference frequency, $\Delta \omega$.
Unlike conventional harmonic driving,
however, this force field also depends on
position because the interference pattern
that drives the sphere takes the form
of a traveling wave.
The driving force experienced by 
the sphere therefore depends
on its position within its trap.
The state dependence of wave-mediated
driving gives rise to
a remarkably rich phenomenology
that includes period-doubling paths
to chaos and Fibonacci cascades 
in addition to
robust scanning modes
that hold the promise of practical applications
\cite{hirayama2019volumetric,fushimi2020ultimate}.

The vertical acoustical trap in our
instrument is created by a 
\SI{40}{\kilo\hertz} piezoelectric
transducer (Sunnytec Electronics, STC-4SH-3540(B))
that is driven near resonance by
a function generator
(Stanford Research Systems DS345).
The transducer has a conical aluminum
horn with a \SI{4.5}{\cm}-diameter
output coupler that slightly concentrates
the sound wave on the acoustical axis.
This wave is reflected by
a horizontal plexiglas sheet
to form a standing-wave acoustical trap.
The plexiglas reflector is positioned
around \SI{6}{\cm} above
the transducer using a micrometer
drive (Prior FB201 Manual Focus Block).
Its height is adjusted to maximize
the amplitude of the pressure
wave at the reflector's surface,
as measured
by a piezoelectric transducer
and a lock-in
amplifier (Stanford Research Systems,
SR830) referenced to the function
generator.

\begin{figure}
    \centering
    \includegraphics[width=0.9\columnwidth]{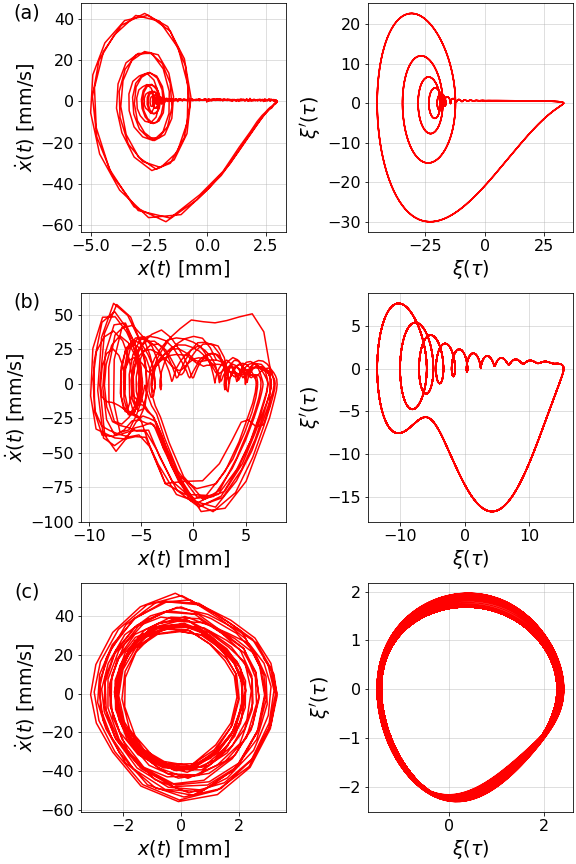}
    \caption{Comparison between experimentally measured
    trajectories, $x(t)$, in the left column 
    and numerical solutions to
    Eq.~\eqref{eq:dimensionlessmodel}, $\xi(\tau)$, in the
    right column. Data are plotted as
    phase space trajectories, $\dot{x}(x)$ and $\xi^\prime(\xi)$,
    respectively.
    (a) $\Delta f = \SI{0.1}{\hertz}$; data replotted
    from Fig.~\ref{fig:schematic}(c). $\Omega = 1/20$,
    $\epsilon = \num{33}.$
    (b) $\Delta f = \SI{1.0}{\hertz}$; $\Omega = 1/3$,
    $\epsilon = \num{15}$.
    (c) $\Delta f = \SI{2.5}{\hertz}$;
    $\Omega = \num{1}$, $\epsilon = \num{1}$.
    All calculations assume $b = \num{0.3}$.
    }
    \label{fig:comparison}
\end{figure}

The transverse traveling wave is
created by a smaller \SI{40}{\kilo\hertz}
piezoelectric transducer (American Piezo, 10-3155) 
driven by a second function generator (Feeltech, FY6600)
whose time base is tied to the first
to ensure relative phase stability.
The frequency and amplitude of this
wave are adjusted to control the
dynamical state of the levitated bead.

The transparent reflector also
provides optical access, allowing us to
record the sphere's motions in the 
horizontal plane
at \SI{40}{frames\per\second}
with a magnification of 
\SI{0.1}{\mm\per pixel}
by a vertically oriented video camera
(FLiR, Flea 3 monochrome).
Figure~\ref{fig:schematic}(b) presents
one video frame
 showing a
\SI{2}{\mm}-diameter styrofoam
bead stably levitated roughly
\SI{1}{\mm} above the transducer's
surface by a
standing wave at $f = \omega/(2\pi) = \SI{40.3}{\kilo\hertz}$.

The trace in Fig.~\ref{fig:schematic}(c)
shows the bead's measured trajectory
along $\hat{x}$ when the driving
wave is detuned from the trapping
wave by $\Delta f = \Delta\omega/(2 \pi) = \SI{0.1}{\hertz}$.
The bead's position in each video
frame is monitored with \SI{10}{\um}
precision using the \texttt{trackpy}
particle-tracking library
\cite{crocker1996methods,dan_allan_2019_3492186}.
This trace embodies both
the promise and the challenge of
the spectral control afforded by
acoustical forces.
Rather than undergoing simple harmonic
motion, the floating bead sweeps out
a nearly linear sawtooth pattern
with distinct ringing on the flyback.
This mode of motion resembles the linear
scanners that have been
demonstrated with acoustical and optical
holographic trapping 
\cite{zheng2012acoustic,ochiai2014three,hirayama2019volumetric}. 
Whereas holographic control requires
hundreds or millions of dynamically controlled
monochromatic wave sources,
spectral control achieves equivalent
motion along a single axis
using just two steady-state waves.

\begin{figure*}
    \centering
    \includegraphics[width=0.9\textwidth]{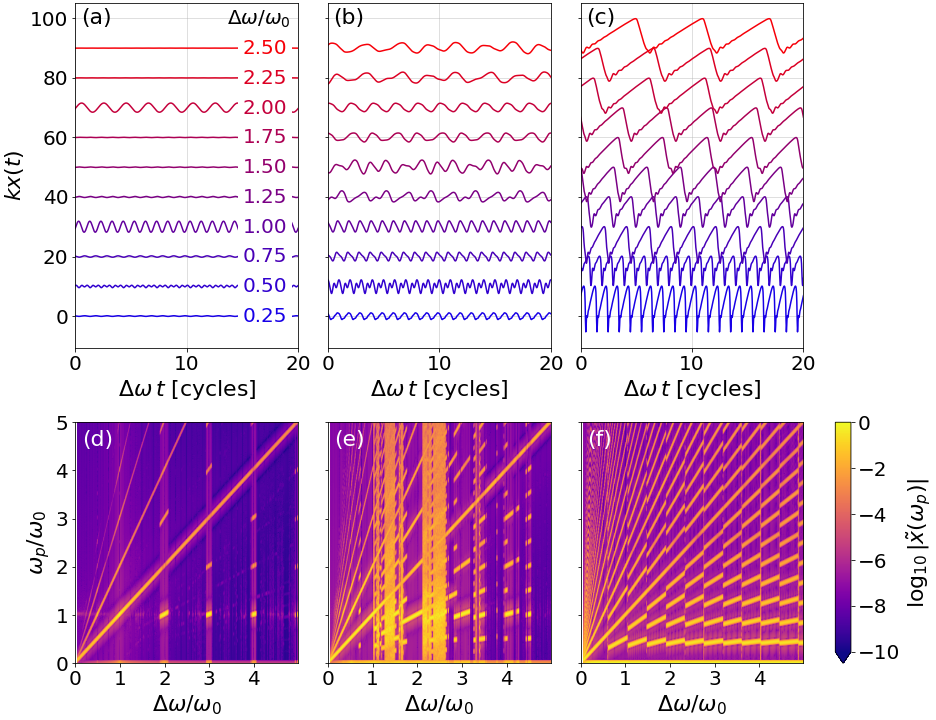}
    \caption{Computed trajectories
    for (a) weak driving ($\epsilon = 0.1$,
    $b = 0.01$), (b) critical driving
    ($\epsilon = 1$, $b = 0.1$), and
    (c) strong driving
    ($\epsilon = 10$, $b = 1$)
    for driving frequencies ranging
    from $\Delta \omega = \SI{0.25}{\omega_0}$
    to $\Delta \omega = \SI{2.5}{\omega_0}$.
    Curves are offset 
    in steps of \num{10} for clarity
    and are colored by values of $\Delta\omega/\omega_0$.
    The corresponding power
    spectra, $\abs{\tilde{x}(\omega_p)}$, 
    in (d), (e) and (f) reveal
    markedly different behavior in the
    different dynamical regimes.
    These spectrograms are computed
    at frequency intervals of
    \SI{0.05}{\omega_0} and
    cover the range from 
    $\Delta \omega = \SI{0.1}{\omega_0}$
    to \SI{5}{\omega_0}.
    Each is normalized
    to its maximum value.
    }
    \label{fig:numericaltrajectories}
\end{figure*}

Sawtooth oscillation 
is just one of many
modes of motion for
the wave-driven acoustical oscillator.
The range of possibilities
is illustrated by an idealized
model for the dynamics
of an object in the acoustical
force field.
Working in Cartesian coordinates with
$\hat{z}$ aligned vertically and
$\hat{x}$ directed along the travelling
wave, the sphere's horizontal displacement
from its equilibrium position may be modelled
as
\begin{equation}
  \label{eq:model}
  \frac{d^2x}{dt^2} + \gamma \frac{dx}{dt} + \omega_0^2 x
  = \frac{F_0}{m} \, e^{i(kx - \Delta \omega t)},
\end{equation}
where $\gamma$ is the damping rate, $k$ is the wavenumber of the
traveling wave,
and $F_0$ is the scale of the transverse
driving force.
This model for the driving force
is derived from the theory of
acoustical forces
in Appendix~\ref{sec:acousticalforces}.
To explore the phenomenology of the wave-driven oscillator, we recast 
Eq.~\eqref{eq:model}
in terms of the dimensionless displacement,
$\xi(\tau) = k x$, and
the dimensionless time, $\tau = \omega_0 t$,
to obtain the non-dimensional equation of motion
\begin{equation}
\label{eq:dimensionlessmodel}
    \xi'' + b \xi' + \xi
    =
    \epsilon \exp(i\xi - i\Omega \tau),
\end{equation}
where $b = \gamma/\omega_0$, 
$\Omega = \Delta \omega/\omega_0$,
$\epsilon = k F_0/(m \omega_0^2)$,
and primes denote derivatives with
respect to $\tau$.

Despite its close resemblance to the
canonical model for a periodically driven
harmonic oscillator,
Eq.~\eqref{eq:dimensionlessmodel}
does not have analytic solutions.
We therefore 
solve Eq.~\eqref{eq:dimensionlessmodel} numerically
using the LSODA integrator with \num{128} steps
per driving period, running the integration
for \num{2048} periods to permit transients to decay before recording trajectory
data for analysis.

Figure~\ref{fig:comparison} provides a sense
of the variety of operating
states of the experimental oscillator and the ability
of Eq.~\eqref{eq:dimensionlessmodel} to capture its behavior.
The left column shows experimentally measured
phase-space trajectories, $\dot{x}(x)$, for
three different values of the detuning:
(a) $\Delta f = \SI{0.1}{\hertz}$, which also
appears in Fig.~\ref{fig:schematic}(c); 
(b) $\Delta f = \SI{1}{\hertz}$; and
(c) $\Delta f = \SI{2.5}{\hertz}$.
The right column shows corresponding 
phase-space trajectories, $\xi^\prime(\xi)$, 
that are computed with Eq.~\eqref{eq:dimensionlessmodel}.
These are not fits to the experimental data,
but rather are comparisons at roughly
equivalent ratios of driving frequencies,
with fixed damping rate, $b = \num{0.3}$,
and with driving amplitude adjusted to
yield good qualitative agreement.
Differences between the experimental and
numerical solutions may be ascribed to nonideal
properties of the physical system, but also 
reflect the immense richness of the system's
full phenomenology.

Figure~\ref{fig:numericaltrajectories}
presents numerical solutions to 
Eq.~\eqref{eq:dimensionlessmodel}
in
three dynamical regimes:
(a) weak driving, 
$\epsilon \ll 1$,
in which the particle undergoes
nearly harmonic oscillations;
(b) critical driving, 
$\epsilon = 1$,
in which harmonic oscillation gives way to chaos; and
(3) strong driving,
$\epsilon \gg 1$,
which is characterized predominantly
by sawtooth 
oscillations.
All three modes of operation are evident
in the experimentally observed behavior
of a levitated sphere.

Figure~\ref{fig:numericaltrajectories} also presents
power spectra,
\begin{equation}
    \abs{ \tilde{x}(\omega_p) }
    =
    \abs{\int_{-\infty}^\infty
    W(t) \, x(t) \, e^{i \omega_p t} \, dt },
\end{equation}
computed from the sphere's trajectory
for each of the three sets of driving amplitudes
and for driving frequencies ranging from $\Delta \omega = \SI{0.1}{\omega_0}$
to $\Delta \omega = \SI{5}{\omega_0}$.
In each case, the dimensionless
damping rate is set to $b = \epsilon/10$.
Power spectra are computed from \num{128} driving cycles
obtained after transients have decayed.
Numerical artifacts due to the finite
signal duration are suppressed with
a normalized Blackman-Harris
window function, $W(t)$.
Driving and response frequencies
in Figs.~\ref{fig:numericaltrajectories}(d), (e) and (f)
are scaled by the oscillator's natural frequency, 
$\omega_0$, which reflects the kinematics of the
sphere in the optical trap, rather than an
inherent property of the sound.

The continuous unit-slope diagonal peaks in all three spectrograms 
represent the oscillator's response at the driving frequency, 
which is the expected behavior for a driven harmonic oscillator.
Features above the diagonal represent harmonics of the driving.
Those below represent subharmonics.

The weakly driven oscillator in Fig.~\ref{fig:numericaltrajectories}(d)
displays a spectrum that is composed of (continuous) harmonics
of the driving frequency superposed with (discrete) harmonics of the
natural frequency, the latter appearing when the driving frequency is an
integer multiple of the natural frequency.
These qualitative features are captured by a
perturbative treatment of Eq~\eqref{eq:dimensionlessmodel}
in Appendix~\ref{sec:weakdriving}.

The spectrum becomes far more complex at critical driving in
Fig.~\ref{fig:numericaltrajectories}(e), with harmonics of the
natural frequency being joined by subharmonics of the natural frequency
and domains of broadband noise.

The remarkable spectrogram in
Fig.~\ref{fig:numericaltrajectories}(f)
suggests that strong driving
supports a sequence of abrupt
transitions between dynamical
states characterized by subharmonics
of the driving frequency.
Each state is composed of sawtooth waves
whose appearance under strong driving conditions
is  motivated by a simplified model in Appendix~\ref{sec:strongdriving}.
Whereas such subharmonics are tied
to resonance-like
harmonics of the fundamental
frequency under weak driving
conditions in Fig.~\ref{fig:numericaltrajectories}(d),
they appear as continuous bands
centered around period-2 
subharmonics of the fundamental
under strong driving conditions.
The transitions between these dynamical
states
are not simply aligned with multiples of
the fundamental frequency,
nor, indeed, are they simply step-like.

\begin{figure}
    \centering
    \includegraphics[width=\columnwidth]{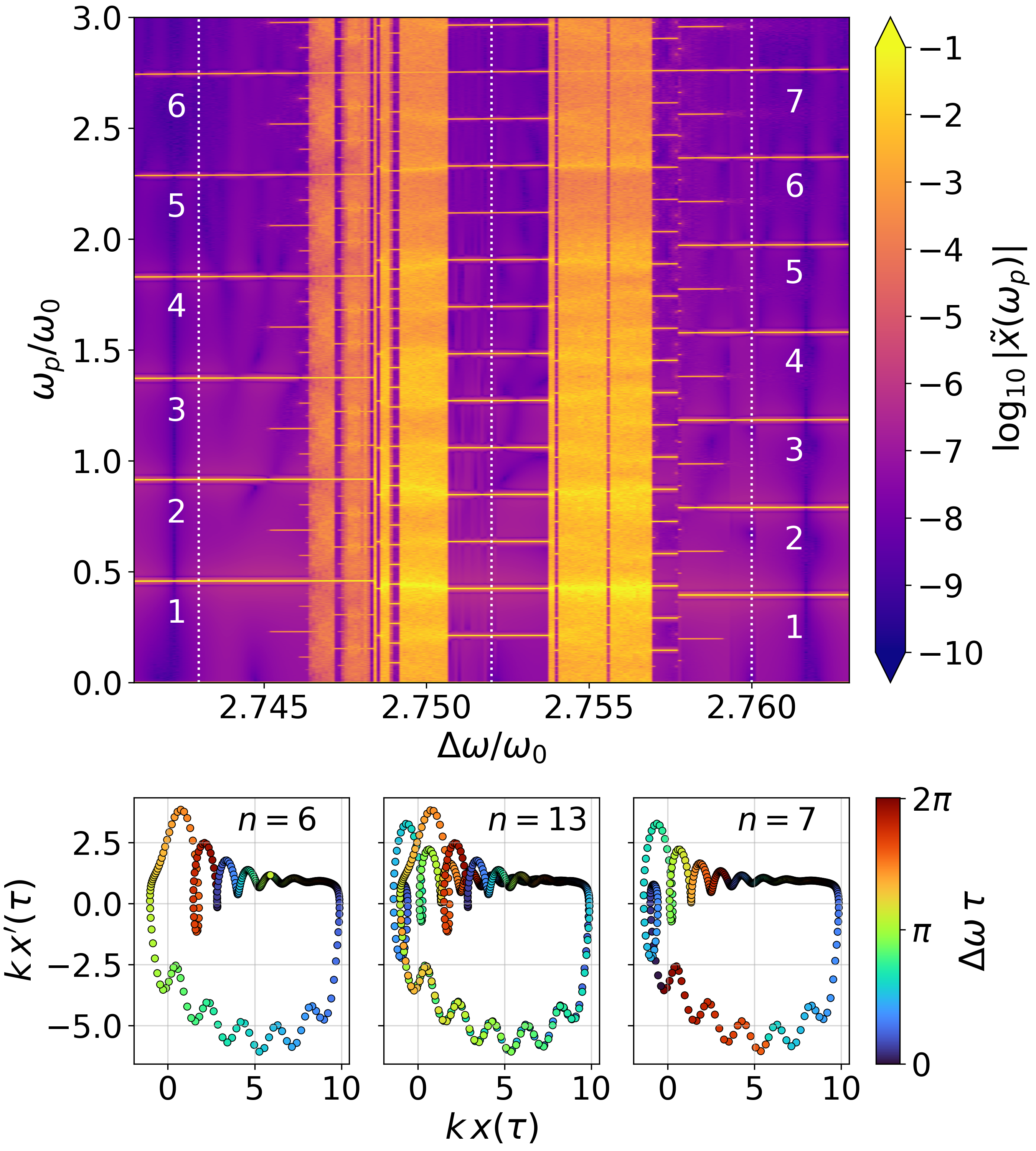}
    \caption{Detailed view of the transition from the $n=6$
    subharmonic state to the $n=7$ state, showing a period-doubling
    path to chaos and a Fibonacci cascade. Phase-space plots show
    one cycle of $n = 6$ and $n = 7$ states bounding the Fibonacci
    $n = 13$ state which appears as a strictly alternating
    sequence of $n = 6$ and $n = 7$ trajectories. Trajectory
    points are colored by periods of the driving frequency
    as indicated in the color bar. Vertical dotted
    lines on the spectrogram indicate driving 
    frequencies for the phase-space trajectories.}
    \label{fig:fibonacci}
\end{figure}

Figure~\ref{fig:fibonacci} presents a higher resolution
map of the transition between period-6 and
period-7 states that appears as a simple step
near $\Omega = \num{2.75}$ in Fig.~\ref{fig:numericaltrajectories}(f).
The continuous peak near the top of this plot
corresponds to the fundamental response,
$\omega_p = \Delta\omega$.
Numbered peaks identify subharmonics
with the family of $n = 6$ subharmonics
on the low-frequency side and
$n = 7$ subharmonics on the high-frequency side.

Following the states up from low frequency
reveals that the $n = 6$ state transitions
to $n = 12$ and $n = 24$, following a period-doubling
route to a chaotic state characterized by a 
broad-band power spectrum.
Intriguingly, the period-doubling cascade
from $n = 7$ toward lower frequencies
appears to be cut off by the incommensurability
between $n = 6$ and $n = 7$ states.

The phase-space plots in Fig.~\ref{fig:fibonacci}
show that the $n = 7$ state principally differs from 
the $n = 6$ state by the appearance of an additional 
loop in the phase-space trajectory.
Vertical (white) dotted lines indicate
the frequencies at which these trajectories were
computed.
Rather than appearing continuously
the additional loop is introduced through a cascade
of intermediate states.
The intermediate $n = 13$ state, for example,
is composed of a strict alternation between
$n = 6$ and $n = 7$ states. The sampled frequency
also is indicated in Fig.~\ref{fig:fibonacci}.
We therefore recognize $n = 13$ as a Fibonacci
state composed of the sum of the two states to either
side in the frequency representation.
Higher-order Fibonacci states also appear
in the transition from $n = 6$ to $n = 7$, involving
more complicated sequences of $n = 6$ and $n = 7$.
Similar combinations of period-doubling cascades to
chaos and Fibonacci cascades appear at the transitions
between each of the principal subharmonic states
in Fig.~\ref{fig:numericaltrajectories}.
Comparable intermediate states appear at each of
the steps in Fig.~\ref{fig:numericaltrajectories}(f).

This study introduces a one-dimensional acoustical
scanner created from two waves. Adding an orthogonal
driving transducer at a third frequency
creates opportunities for two-dimensional scanning.
More generally, three detuned traveling
waves should create a dynamic volumetric
force field that can translate one or more 
object through three-dimensional trajectories
that are programmed by the differences in the
waves' frequencies. Dynamic control over those
frequencies promises still more dextrous
control in three dimensions while using fewer transducers
than a holographic array.
Faster motion in this case is achieved with larger
detuning, which is technically simpler than
rapid reconfiguration of acoustic holograms.

Realizing the potential of spectral control 
for practical large-scale manipulation
requires a deeper understanding
of the dynamics of wave-driven harmonic oscillators.
Analytic solutions to Eq.~\eqref{eq:dimensionlessmodel},
for the minimal example presented in this study,
for example, would guide development of
practical acoustomechanical scanners.
Detuned coaxial acoustical Bessel beams, for example,
should act as high-speed tractor beams for
transporting material \cite{marston2006axial}
through their analogy to optical conveyors
\cite{cizmar2005optical,ruffner2012optical}.
Spectral control of acoustical force landscapes \cite{abdelaziz20}
also will create new opportunities for influencing
the organization of multi-particle systems
\cite{lim2019cluster} by introducing a new
mechanism for controlling inter-particle
interactions.

This work was supported by the MRSEC program of
the National Science Foundation under
Award Number DMR-1420073.

\appendix

\renewcommand{\theequation}{A\arabic{equation}}

\section{Acoustical forces in the wave-driven oscillator}
\label{sec:acousticalforces}

The time-averaged acoustic radiation force on a small sphere 
of radius $a_p$ at position $\vec{r}$ in a harmonic
pressure field $p(\vec{r}, t)$ 
is well-approximated by the Gor'kov potential \cite{gorkov1962forces,bruus2012acoustofluidics}:
\begin{equation}
    \label{eq:gorkov}
    U(\vec{r}) 
    = \pi a_p^3 \rho_m \left[ 
    f_0 \frac{\avg{\abs{p(\vec{r},t)}^2}}{3 \rho_m^2 c_m^2} - 
    f_1 \frac{\avg{\abs{\vec{v}(\vec{r},t)}^2}}{2}
    \right],
\end{equation}
where the coupling due to compressibility and density mismatches between the particle and medium are respectively described by
\begin{subequations}
\begin{align}
    f_0 &= 1-\frac{\rho_m c_m^2}{\rho_p c_p^2} \quad \text{and} \\
    f_1 &= \frac{2(\rho_p-\rho_m)}{2\rho_p+\rho_m} .
\end{align}
\end{subequations}
The symbols $\rho$ and $c$ denote density and sound speed, respectively, 
Subscripts $p$ and $m$ refer to the particle and medium, respectively.
In the system under study, the acoustic contrast between the sphere and air
is large enough that $f_0 \approx f_1 \approx 1$.
Angle brackets in Eq.~\eqref{eq:gorkov} signify time averages.

The velocity field is related to the pressure field in linear
sound by Newton's second law,
\begin{equation}
    \rho_m \frac{\partial \vec{v}}{\partial t} 
    + \nabla p = 0.
\end{equation}
For a harmonic wave at frequency $\omega$, therefore,
the pressure acts as a scalar potential
for the velocity,
\begin{equation}
    \label{eq:velocity}
    \vec{v}(\vec{r},t) 
    =
    - \frac{i}{\omega \rho} \nabla p.
\end{equation}
Equations~\eqref{eq:gorkov} and \eqref{eq:velocity} define
the acoustic force field established by 
a specified pressure field, $p(\vec{r},t)$.

The wave-driven oscillator is subject to two harmonic pressure fields:
a standing wave, 
\begin{subequations}
\begin{equation}
    \label{eq:standingwave}
    p_1(\vec{r}, t) 
    = 
    u_1 \sin(k_1 z) \, e^{-i \omega_1 t}
\end{equation}
at frequency $\omega_1$ that forms an acoustical trap
and a traveling wave, 
\begin{equation}
    \label{eq:travelingwave}
    p_2(\vec{r}, t) 
    = 
    u_2 \, e^{i k_2 x} \, e^{-i \omega_2 t},
\end{equation}
at frequency $\omega_2$
that provides the driving.
\end{subequations}
To ensure that the amplitudes $u_1$ and $u_2$ are real-valued
we absorb spatial and temporal phase differences into
the definitions of $x$ and $t$.

The total pressure field, $p(\vec{r}, t) = p_1(\vec{r},t) + p_2(\vec{r},t)$,
has an intensity
\begin{equation}
    \abs{p(\vec{r},t)}^2
    = 
    \abs{p_1}^2 + \abs{p_2}^2
    + 2 \Re\{p_1^\ast p_2\}
\end{equation}
that includes time-independent single-field intensities
\begin{equation}
    \abs{p_1}^2 + \abs{p_2}^2
    =
    u_1^2 \sin^2(k_1 z) + u_2^2
\end{equation}
and a time-dependent cross term,
\begin{multline}
    \label{eq:cross}
    2 \Re\{p_1^\ast p_2\}
    =
    2 u_1 u_2 \sinc(\phi) \times \\ 
    \sin(k_1 z) \, \cos(k_2 x - \Delta\omega t - \phi),
\end{multline}
that oscillates at the difference
frequency, $\Delta \omega = \omega_2 - \omega_1$.
The phase offset, $\phi = \pi \Delta \omega/\omega$,
depends on the detuning of the two waves relative
to the center frequency $\omega = (\omega_1 + \omega_2)/2$
and contributes to the amplitude of the cross term.
Pursuing a similar line of reasoning for the
velocity yields
\begin{align}
    \abs{\vec{v}(\vec{r},t)}^2
    & = 
    \abs{\vec{v}_1}^2 + \abs{\vec{v}_2}^2
    +
    2 \Re\{\vec{v}_1^\ast \cdot \vec{v}_2 \} \\
    & =
    \frac{u_1^2}{\rho_m^2 c_m^2} \cos^2(k_1 z)
    + 
    \frac{u_2^2}{\rho_m^2 c_m^2},
\end{align}
with the cross term vanishing because the two
waves are orthogonal.
Dropping constant terms, we obtain a time-dependent
potential,
\begin{equation}
    U(\vec{r}, t)
    =
    A
    \cos^2(k_1 z) +
    B \sin(k_1 z)
    \cos(k_2 x - \Delta\omega t),
\end{equation}
that includes an axial trapping potential
whose scale is set by the standing wave,
\begin{equation}
    A = 
    \frac{1}{6} u_1^2 
    \frac{\pi a_p^3}{\rho_m c_m^2},
\end{equation}
and a time-dependent contribution
that arises from the interference between
the standing and traveling wave,
\begin{equation}
    B = 
    \frac{2}{3} u_1 u_2 
    \frac{\pi a_p^3}{\rho_m c_m^2}.
\end{equation}

In the absence of external forces, the
trapping potential due to the standing wave
would tend to localize the particle
near $k_1 z = \pi/2$, which would
effectively suppress the transverse
driving force.
Gravity, however, displaces the particle
downward to
\begin{equation}
    z_0 \approx 
    - \frac{1}{2k_1} 
    \sin\left(
    8 \frac{\rho_p \rho_m c_m^2 g}{u_1^2}
    \right),
\end{equation}
where $g$ is the acceleration due to gravity.
The resulting 
transverse force then becomes
\begin{equation}
\label{eq:drivingforce}
    F(\vec{r}, t)
    =
    F_0 \,
    \sin(k_2 x - \Delta\omega t),
\end{equation}
with the force scale
$F_0 = -k_2 B \sin(k_1 z_0)$.
This motivates the choice of 
a wavelike driving
force for the right-hand
side of Eq.~\eqref{eq:model}
in the main text.

In the interest of clarity, this analysis has
not accounted for the transverse
trapping force arising from transverse
gradients in the
vertical standing wave, but rather has
treated the standing wave as
plane-like near the trapping plane.
A more complete treatment yields
an expression comparable to Eq.~\eqref{eq:drivingforce}
with corrections to the scale factor, $F_0$.
Interestingly, wave-like driving in this
system appears to rely on gravity to displace
the trapped particle.
We predict therefore that the effects
observed in this study would not occur
under density-matched conditions
or in microgravity.

\section{Weak driving in the underdamped regime}
\label{sec:weakdriving}

The wave-driven oscillator does not
behave like a standard periodically-driven
harmonic oscillator even when weakly
driven, as illustrated by Fig.~\ref{fig:numericaltrajectories}(d)
in the main text.
We explain qualitative features of
this system's dynamics
by treating Eq.~\eqref{eq:dimensionlessmodel}
from the main text
perturbatively in the limit of weak driving
($\epsilon < 1$) when damping
may be ignored ($b \ll 1$):
\begin{equation}
    \xi''(\tau) + \xi(\tau) 
    \approx 
    \epsilon 
    e^{i\xi(\tau) - i\Omega\tau }.
\end{equation}
Substituting a trial solution
\begin{equation}
    \xi(\tau) 
    = 
    \xi_0(\tau)
    + \epsilon \xi_1(\tau) 
    + \epsilon^2 \xi_2(\tau)+\ldots,
\end{equation}
with initial conditions 
$\xi_n(0) = \xi_n^\prime(0)=0$ 
yields for the first few terms
\begin{subequations}
\begin{align}
    \xi_0(\tau) & = 0 
    \label{eq:xi0} \\
    \xi_1(\tau)
    & =
    \frac{e^{-i\Omega\tau}}{1-\Omega^2}
    +
    \frac{1}{2}
    \left(
    \frac{e^{-i\tau}}{\Omega-1} -
    \frac{e^{i\tau}}{\Omega+1}
    \right)
    \label{eq:xi1} \\
    \xi_2(\tau)
    & =
    i \frac{e^{-2i\Omega\tau}}{(\Omega^2-1)(4\Omega^2-1)} 
    \nonumber \\
    &- \frac{i}{2 \Omega} 
        \left[
        \frac{e^{-i \tau}}{(\Omega-2)(2\Omega-1)} - 
        \frac{e^{i\tau}}{(\Omega+2)(2\Omega+1)}
        \right]
        \nonumber \\
    &- \frac{i}{2\Omega}
        \left[
        \frac{e^{-i (\Omega+1)\tau}}{(\Omega-1)(\Omega+2)} -
        \frac{e^{-i(\Omega-1)\tau}}{(\Omega+1)(\Omega-2)}
        \right] .
    \label{eq:xi2}
\end{align}
\end{subequations}
The first-order contribution 
in Eq.~\eqref{eq:xi1}
includes a response at the the driving
frequency, $\Omega$, that is expected for a driven
harmonic oscillator.
It also includes a response
at the natural frequency, $\omega_p = \omega_0$, 
that appears regardless
of the driving frequency, and is resonantly
enhanced when $\Omega = \pm 1$.
This appears in Fig.~\ref{fig:numericaltrajectories}(d)
as a horizontal streak at $\omega_p = \omega_0$.

The second-order contribution
from Eq.~\eqref{eq:xi2} includes the second harmonic
of the driving frequency, which appears in 
Fig.~\ref{fig:numericaltrajectories}(d) as a continuous
diagonal streak at $\omega_p = 2 \Delta \omega$.
The second term in Eq.~\eqref{eq:xi2} contributes
to the oscillator's response at the natural
frequency, with additional resonant enhancement
at $\Omega = \pm 2$
and $\Omega = \pm 1/2$.
The third term describes sum- and difference-frequency
responses at
$\omega_p = \Delta \omega \pm \omega_0$
that are resonantly enhanced near 
$\Omega = 2$.
This contribution accounts for the frequency dependence
of the observed peaks in Fig.~\ref{fig:numericaltrajectories}(d)
near $\omega_p = \omega_0$
and $\omega_p = 3 \omega_0$ 
for driving frequencies
in the vicinity of $\Delta \omega = 2 \omega_0$.

Higher-order terms introduce corresponding harmonics
of the driving frequency as well as the sum- and
difference-frequency responses that account for the
grid of discrete resonant peaks in
Fig.~\ref{fig:numericaltrajectories}(d).

\section{Strong driving in the overdamped regime}
\label{sec:strongdriving}

The sawtooth oscillations observed
in Fig.~\ref{fig:numericaltrajectories}(c) from the main text
under strong driving conditions, 
$\epsilon > 1$, feature long
stretches of nearly linear motion punctuated by abrupt changes in
direction.
The equation of motion
from Eq.~\eqref{eq:dimensionlessmodel} in the main text
may be approximated in these
weakly inertial passages
by the transcendental
equation
\begin{equation}
    \label{eq:strongdriving}
    \xi = \epsilon \sin(\xi - \Omega \tau).
\end{equation}
For large $\epsilon$, the object follows the wave
until $\xi \approx \epsilon$, at which point the
restoring force overcomes the driving force
and the object snaps back.
This occurs at $\tau = \epsilon/\Omega$
which suggests the the fundamental oscillation
frequency,
\begin{equation}
    \omega_p = 2 \pi \Delta\omega \, \frac{m \omega_0^2}{k F_0},
\end{equation}
can be substantially lower than the
driving frequency, $\Delta \omega$ because
the wave may have to advance by several wavelengths
before the object can break free.

Once the object is free, it travels from
$\xi = \epsilon$ back toward $\xi = 0$
until it once again encounters a region
of the traveling wave that can stop it
and reverse its motion.
The sawtooth oscillator's recovery is not
captured by Eq.~\eqref{eq:strongdriving}
and requires a more complete treatment of
the sphere's dynamics.
Although this model motivates the observed
mode of oscillation, therefore, it does
not capture such properties as the
transition points between operating modes
or the complicated dynamical behavior
near transitions.

%\bibliographystyle{publist}
%\bibliography{oscillations}

\end{document}